\documentclass[reprint,aps,prd,showpacs]{revtex4-1}
\usepackage{graphicx}
\usepackage{amsfonts}
\usepackage{amsmath,amsthm,amssymb}
\newtheorem{theorem}{Theorem}
\begin{document}

\title{New Approach to Nonperturbative Quantum Mechanics with Minimal Length Uncertainty}
\author{Pouria Pedram}
\email{p.pedram@srbiau.ac.ir}
\affiliation{Department of Physics,
Science and Research Branch,\\ Islamic Azad University, Tehran,
Iran}

\date{\today}

\begin{abstract}
The existence of a minimal measurable length is a common feature of
various approaches to quantum gravity such as string theory, loop
quantum gravity and black-hole physics. In this scenario, all
commutation relations are modified and the Heisenberg uncertainty
principle is changed to the so-called Generalized (Gravitational)
Uncertainty Principle (GUP). Here, we present a one-dimensional
nonperturbative approach to quantum mechanics with minimal length
uncertainty relation which implies $X=x$ to all orders and
$P=p+\frac{1}{3} \beta p^3$ to first order of GUP parameter $\beta$,
where $X$ and $P$ are the generalized position and momentum
operators and $[x,p]=i\hbar$. We show that this formalism is an
equivalent representation of the seminal proposal by Kempf, Mangano,
and Mann and predicts the same physics. However, this proposal
reveals many significant aspects of the generalized uncertainty
principle in a simple and comprehensive form and the existence of a
maximal canonical momentum is manifest through this representation.
The problems of the free particle and the harmonic oscillator are
exactly solved in this GUP framework and the effects of GUP on the
thermodynamics of these systems are also presented. Although $X$,
$P$, and the Hamiltonian of the harmonic oscillator all are formally
self-adjoint, the careful study of the domains of these operators
shows that only the momentum operator remains self-adjoint in the
presence of the minimal length uncertainty. We finally discuss the
difficulties with the definition of potentials with infinitely sharp
boundaries.
\end{abstract}

\pacs{04.60.Bc}

\keywords{Quantum gravity; Generalized uncertainty principle;
Minimal length uncertainty relation.}

\maketitle

\section{Introduction}
The unification of general relativity with the laws of quantum
mechanics is one of the oldest wishes of theoretical physicists from
the birth of quantum mechanics. We can mention the canonical
quantization \cite{Dewitt} and the path integral quantization of
gravity \cite{Hawking} as two well-known but old proposals which
tried to present a quantization scheme for gravity. However, from
the field theoretical viewpoint, the theory of relativity is not
renormalizable and leads to ultraviolet divergencies. Moreover,
around the Planck energy scale, the effects of gravity are so
important that they would result in discreteness of the spacetime
manifold. This argument is based on the fact that, when we try to
probe small distances with high energies, it will significantly
disturb the spacetime structure by the gravitational effects.
However, the theory can be renormalizable by introducing a minimal
observable length as an effective cutoff in the ultraviolet domain.

The existence of a minimum measurable length is one of the common
aspects of various candidates of quantum gravity such as string
theory, loop quantum gravity, and quantum geometry. Within a
string-theoretical argument, we can say that a string cannot probe
distances smaller than its length. Moreover, some Gedanken
experiments in black-hole physics and noncommutativity of the
spacetime manifold all agree on the existence of a minimal
observable distance of the order of the Planck length
$\ell_{Pl}=\sqrt{G\hbar/c^3}\approx 10^{-35}m$, where $G$ is
Newton's constant \cite{r1,r2,r3,r4,r5}. In fact, the finite
resolution of spacetime points is a consequence of finite time
measurement. In principle, one can probe very short distances in
$D0$-branes but in an infinite time.

Note that, this is in obvious contradiction with the Heisenberg
Uncertainty Principle (HUP) which puts no lower or upper bound on
the nonsimultaneous measurement of the position or the momentum of a
particle. In fact, in ordinary quantum mechanics $\Delta X$ can be
made arbitrarily small by letting $\Delta P$ to grow
correspondingly. However, for energies close to the Planck energy,
the particle's Schwarzschild radius and its Compton wavelength
become approximately in the order of the Planck length. So, in order
to merge the idea of the minimal length into quantum mechanics, we
need to modify the ordinary uncertainty principle to the so-called
Generalized Uncertainty Principle (GUP). Indeed, the notion of
minimal length should quantum mechanically be described as a minimal
uncertainty in position measurements. The introduction of this idea
has drawn much attention in recent years and many papers have been
appeared in the literature to address the effects of GUP on various
quantum mechanical systems and phenomena
\cite{12,13,14,15,16,17,18,18-2,19,101,102,103,104,105,106,pedram2,NP}.

In this paper, we present a nonperturbative approach to
one-dimensional gravitational quantum mechanics which implies a
minimal length uncertainty so that the generalized position operator
does not change to all orders, that is, $X=x$ and the generalized
momentum operator is given by $P=p+ \frac{1}{3}\beta p^3$ to first
order of the GUP parameter. In this formalism the generalized
position and momentum operators satisfy $[X,P]=i\hbar(1+\beta
P^{2})$ where $x$ and $p$ are the ordinary position and momentum
operators $[x,p]=i\hbar$. We show that this proposal is equivalent
with Kempf, Mangano, and Mann (KMM) representation, but it only
modifies the kinetic part of the Hamiltonian and has no effect on
the potential part. Moreover, this representation agrees with
perturbative approaches and predicts the presence of a maximal
canonical momentum $p_{max}$. Here, we consider the problems of the
free particle and the harmonic oscillator in the context of the
generalized uncertainty principle and obtain the exact eigenvalues
and corresponding eigenfunctions. Then, we discuss the consequences
of the minimal uncertainty in position measurement on the partition
function, mean energy, and heat capacity of these systems. The
difficulties with potentials with infinitely sharp boundaries are
also presented.

\section{The Generalized Uncertainty Principle}
According to the Heisenberg uncertainty relation, in principle, we
can separately measure the position and momentum of particles with
arbitrary precision. Thus, if there is a genuine lower bound on the
results of the measurements, the Heisenberg uncertainty relation
should be modified. Here we consider a generalized uncertainty
principle which results in a minimum observable length
\begin{eqnarray}\label{gup}
\Delta X \Delta P \geq \frac{\hbar}{2} \left( 1 +\beta (\Delta P)^2
+\zeta \right),
\end{eqnarray}
where $\beta$ is the GUP parameter and $\zeta$ is a positive
constant that depends on the expectation value of the momentum
operator. We also have $\beta=\beta_0/(M_{Pl} c)^2$ where $M_{Pl}$
is the Planck mass and $\beta_0$ is of the order of one. Note that
the deviation from the Heisenberg picture takes place in the high
energy limit where the quantum gravity effects are dominant. So, for
the energies much smaller than the Planck energy $M_{Pl} c^2\sim
10^{19}$ GeV, we should recover the famous Heisenberg uncertainty
relation. It is straightforward to check that the above inequality
relation (\ref{gup}) implies the existence of an absolute minimal
length uncertainty as $(\Delta X)_{min}=\hbar\sqrt{\beta}$. In the
context of string theory, we can interpret this length as the string
length. Accordingly, the string's length is proportional to the
square root of the GUP parameter. In one-dimension, the above
uncertainty relation can be obtained from a deformed commutation
relation, namely
\begin{eqnarray}\label{gupc}
[X,P]=i\hbar(1+\beta P^2),
\end{eqnarray}
where, for $\beta=0$ we recover the well-known commutation relation
in ordinary quantum mechanics. Now using Eqs.~(\ref{gup}) and
(\ref{gupc}) we can find the relation between $\zeta$ and the
expectation value of the momentum operator i.e. $\zeta=\beta\langle
P\rangle^2$. As Kempf, Mangano, and Mann have suggested in their
seminal paper, in momentum space representation, we can write $X$
and $P$ as \cite{Kempf}
\begin{eqnarray}\label{k1}
P \phi(p)&=& p\phi(p),\\ X\phi(p) &=& i\hbar \left( 1 + \beta
p^2\right)\partial_p\phi(p),\label{k2}
\end{eqnarray}
where $X$ and $P$ are symmetric operators on the dense domain
$S_{\infty}$ with respect to the following scalar product
\begin{eqnarray}\label{scalar}
\langle\psi|\phi\rangle=\int_{-\infty}^{+\infty}\frac{\mathrm{d}p}{1+\beta
p^2}\psi^{*}(p)\phi(p),
\end{eqnarray}
where $\int_{-\infty}^{+\infty}\frac{\mathrm{d}p}{1+\beta
p^2}|p\rangle\langle p|=1$ and $\langle p|p'\rangle=\left( 1 + \beta
p^2\right)\delta(p-p')$. In this representation the position
operator is merely symmetric, but $P$ is self-adjoint \cite{Kempf}.
With this definition, the commutation relation (\ref{gupc}) is
exactly satisfied. Also, in quasiposition representation this
formulation results in \cite{Kempf}
\begin{eqnarray}\label{k3}
P \psi(x)&=& \frac{\tan\left(-i\hbar\sqrt{\beta}\partial_x\right)}{\sqrt{\beta}}\psi(x),\\
X\psi(x) &=& \left(x+\beta
\frac{\tan\left(-i\hbar\sqrt{\beta}\partial_x\right)}{\sqrt{\beta}}\right)\psi(x).\label{k4}
\end{eqnarray}
Note that, for the general potential, expressing the position
operator as a combination of ordinary position and momentum
operators results in a complicated high-order generalized
Schr\"odinger equation. So, finding the solutions even for the
simple potentials would not be an easy task.

To overcome this problem, we propose the following
generalized position and momentum operators
\begin{eqnarray}\label{x0p01}
X &=& x,\\ P &=& \frac{\tan\left(\sqrt{\beta}p\right)}{\sqrt{\beta}},\label{x0p02}
\end{eqnarray}
where $x$ and $p$ obey the canonical commutation relation
$[x,p]=i\hbar$. $X$ and $P$ are symmetric operators on the dense
domain $S_{\infty}$ of functions decaying faster than any power
\begin{eqnarray}
\hspace{-1cm}(\langle\psi|X)|\phi\rangle=\langle\psi|(X|\phi\rangle)\hspace{.2cm}\mbox{and}\hspace{.2cm}
(\langle\psi|P)|\phi\rangle=\langle\psi|(P|\phi\rangle),
\end{eqnarray}
but now with respect to the scalar product:
\begin{eqnarray}\label{scalar-product}
\langle\psi|\phi\rangle=\int_{-\frac{\pi}{2\sqrt{\beta}}}^{+\frac{\pi}{2\sqrt{\beta}}}dp\,\psi^{*}(p)
\phi(p),
\end{eqnarray}
The symmetry of $P$ (\ref{x0p02})  is obvious. The symmetry of  $X$
(\ref{x0p01}) can be seen by performing partial integrations
\begin{eqnarray}
&&\int_{-\frac{\pi}{2\sqrt{\beta}}}^{+\frac{\pi}{2\sqrt{\beta}}}dp\,\psi^{*}(p)
\left(i\hbar\frac{\partial}{\partial
p}\right)\phi(p)\nonumber\\
&&=\int_{-\frac{\pi}{2\sqrt{\beta}}}^{+\frac{\pi}{2\sqrt{\beta}}}\mathrm{d}p\,\left(i\hbar\frac
{\partial\psi(p)}{\partial p}\right)^{*}\phi(p),
\end{eqnarray}
which is valid for the functions vanishing at $\pm
\frac{\pi}{2\sqrt{\beta}}$. Indeed, the symmetry property of the
position and momentum operators ensures that all expectation values
are real. This definition exactly satisfies the condition
$[X,P]=i\hbar(1+\beta P^2)$ and agrees with the well-known relations
\cite{main}, namely
\begin{eqnarray}\label{xp-old1}
X &=& x,\\ P &=& p\left( 1 + \frac{1}{3}\beta\, p^2 \right),\label{xp-old2}
\end{eqnarray}
to the first order of the GUP parameter. Note that to
${\cal{O}}(\beta)$, the definitions (\ref{k3}) and (\ref{k4}) result
in $X=x+\beta p$ and $P=p\left( 1 + \frac{1}{3}\beta\, p^2 \right)$
which differ with (\ref{xp-old1}) and (\ref{xp-old2}).

Now, we show that our proposal and KMM representation are equivalent
in essence. Indeed, they are related by the following canonical
transformation:
\begin{eqnarray}
X&\rightarrow&\left[1+\arctan^2\left(\sqrt{\beta}P\right)\right]X,\\ P&\rightarrow&\arctan\left(\sqrt{\beta}P\right)/\sqrt{\beta},
\end{eqnarray}
which transforms (\ref{x0p01}) and (\ref{x0p02}) into (\ref{k1}) and
(\ref{k2}) subjected to condition (\ref{gupc}). We can interpret $P$
and $p$ as follows: $p$ is the momentum operator at low energies
($p=-i\hbar\partial/\partial{x}$) while $P$ is the momentum operator
at high energies. Obviously, this procedure affects all Hamiltonians
in adopted quantum mechanics.

Note that for an operator $A$ which is ``formally'' self-adjoint
($A=A^{\dagger}$) such as (\ref{x0p01}) and (\ref{x0p02}), this does
not prove that $A$ is truly self-adjoint because in general the
domains ${\cal D}(A)$ and ${\cal D}(A^{\dagger})$ may be different.
The operator $A$ with dense domain ${\cal D}(A)$ is said to be
self-adjoint if ${\cal D}(A) ={\cal D}(A^{\dagger})$ and
$A=A^{\dagger}$. For instance, similar to KMM representation, $X$ is
merely symmetric but not self-adjoint. To see this note that in this
representation and in the momentum space the wave function $\phi(p)$
have to vanish at the end of the $p$ interval
$(-\pi/2\sqrt{\beta}<p<\pi/2\sqrt{\beta})$, because the tangent
function diverges there. So, $X$ is a derivative operator
$i\hbar\partial/\partial{p}$ on an interval with Dirichlet boundary
conditions. But this means that $X$ cannot be self-adjoint because
all candidates for the eigenfunctions of $X$, (the plane waves,
which are even normalizable) are not in the domain of $X$ because
they do not obey Dirichlet boundary conditions. Calculating the
domain of the adjoint of $X$ shows that it is larger than that of
$X$, so $X$ is indeed not self-adjoint i.e.
\begin{eqnarray}
&&\int_{-\frac{\pi}{2\sqrt{\beta}}}^{+\frac{\pi}{2\sqrt{\beta}}}\mathrm{d}p\,\psi^{*}(p)\left(i\hbar\frac{\partial}{\partial
p}\right)\phi(p)\nonumber\\ &&=
\int_{-\frac{\pi}{2\sqrt{\beta}}}^{+\frac{\pi}{2\sqrt{\beta}}}\mathrm{d}p\,\left(i\hbar\frac{\partial\psi(p)}{\partial
p}\right)^{*}\phi(p)\nonumber\\
&&\quad+i\hbar\,\psi^*(p)\phi(p)\Bigg|_{p=+\frac{\pi}{2\sqrt{\beta}}}\hspace{-.5cm}-i\hbar\,\psi^*(p)\phi(p)\Bigg|_{p=-\frac{\pi}{2\sqrt{\beta}}}\hspace{-.75cm}.
\end{eqnarray}
Now since $\phi(p)$ vanishes at $p=\pm\frac{\pi}{2\sqrt{\beta}}$,
$\psi^*(p)$ can take any arbitrary value at the boundaries. The
above equation implies that $X$ is symmetric, but it is not a
self-adjoint operator. Although its adjoint
$X^{\dagger}=i\hbar\partial/\partial p$ has the same formal
expression, it acts on a different space of functions, namely
\begin{eqnarray}
{\cal D}(X)&=&\bigg\{\phi,\phi'\in{\cal
L}^2\left(\frac{-\pi}{2\sqrt{\beta}},\frac{+\pi}{2\sqrt{\beta}}\right)\,;
\phi\left(\frac{+\pi}{2\sqrt{\beta}}\right)\nonumber\\ &&\quad\quad=\phi\left(\frac{-\pi}{2\sqrt{\beta}}\right)=0\bigg\},\\
{\cal D}(X^{\dagger})&=&\bigg\{\psi,\psi'\in{\cal
L}^2\left(\frac{-\pi}{2\sqrt{\beta}},\frac{+\pi}{2\sqrt{\beta}}\right);\nonumber\\
&&\quad\quad\mbox{no other restriction on }\psi\bigg\}.
\end{eqnarray}
As it is also shown in Ref.~\cite{Kempf2}, any operator $X$ which
obeys the uncertainty relation (\ref{gup}) is merely symmetric. On
the other hand, since there are no Dirichlet boundary conditions on
the wave functions in the position space ($-\infty<x<\infty$), $P$
is still self-adjoint. In the next section and after finding the
momentum eigenfunctions, we prove the self-adjointness property of
$P$ using von Neumann's theorem.

To proceed further, let us consider the following Hamiltonian
\begin{eqnarray}
H=\frac{P^2}{2m} + V(X),
\end{eqnarray}
which using Eqs.~(\ref{x0p01}) and (\ref{x0p02}) can be written
exactly and also perturbatively as
\begin{eqnarray}\label{H0}
&&H= \frac{\tan^2\left(\sqrt{\beta}p\right)}{2\beta m}+V(x), \\
&&=H_0+ \sum_{n=3}^\infty \frac{(-1)^{n-1} 2^{2n} (2^{2n}-1)(2n-1)
B_{2n}}{2m(2n)!} \beta^{n-2} p^{2(n-1)},\hspace{.5cm}\label{H}
\end{eqnarray}
where $H_0=p^2/2m + V(x)$ and $B_n$ is the $n$th Bernoulli number.
So the corrected terms in the modified Hamiltonian are only momentum
dependent and proportional to $p^{2(n-1)}$ for $n\geq 3$. As we
shall explicitly show, the presence of these terms leads to a
positive shift in the particle's energy spectrum. Note that, in
general, even for the self-adjoint position and momentum operators,
it is by no means obvious that the resulting Hamiltonian will be
self-adjoint until and unless the potential term is specified and
the appropriate domain is chosen. It is worth to mention that all
our calculations are in one-dimensional space. Indeed, in higher
dimensions it is necessary to have noncommutativity of coordinates
in order to satisfy the Jacobi identity as done by KMM \cite{Kempf}.
In one-dimensional space, the Jacobi identity is automatically
satisfied. Also, one may relax the point size property of the
particle as in the string theory. So we can interpret Eq.~(\ref{H0})
as the Schr\"odinger equation for the particle with size $\sim\hbar
\sqrt{\beta}$, where the effect of the nonzero size effectively
appears in the kinetic part of the Hamiltonian.

In the quantum domain, this Hamiltonian results in the following
generalized Schr\"odinger equation in the quasiposition
representation:
\begin{eqnarray}
&-&\frac{\hbar^2}{2m}\frac{\partial^2\psi(x)}{\partial x^2}+
\sum_{n=3}^\infty \alpha_n \hbar^{2(n-1)}\beta^{n-2}
\frac{\partial^{2(n-1)}\psi(x)}{\partial x^{2(n-1)}}\nonumber\\
&&+V(x)\psi(x)=E\,\psi(x),
\end{eqnarray}
where $\alpha_n=2^{2n} (2^{2n}-1)(2n-1) B_{2n}/2m(2n)!$  and the
second term is due to the GUP corrected terms in (\ref{H}).

Among infinite possible canonical transformations (CTs),  our
proposal (\ref{x0p01}) and (\ref{x0p02}) has some useful and novel
properties. First, it does not change the nature of the position
operator and, consequently, the potential term and only modifies the
momentum or the kinetic operator. So, among several CTs, only this
one preserves the ordinary nature of the position operator. Second,
this formalism lets us to write the Hamiltonian as $H=H_0+\beta
H_1+\beta^2 H_2+...$, where $H_0=p^2/2m+V(x)$ is the ordinary
Hamiltonian and $H_1,H_2,...$ contain only the momentum operator.
So, using the perturbation theory, the unperturbed eigenfunctions
satisfy $H_0|\psi_0\rangle=E_0|\psi_0\rangle$ and we can find
$\langle H_1\rangle,\langle H_2\rangle,...$ in an straightforward
manner as done for various cases such as Ref.~\cite{main}. In other
CTs like the KMM proposal, we cannot decompose the Hamiltonian in
such configurations. So, in this sense, this proposal is compatible
with perturbative representations. Third, this proposal predicts the
existence of a maximal canonical momentum. In fact, the particular
form of the kinetic part of the Hamiltonian (\ref{H0}) implies the
existence of a maximal momentum:
\begin{eqnarray}
p_{max}=\frac{\pi}{2\sqrt{\beta}}=\frac{\pi M_{Pl} c}{2\sqrt{\beta_0}},
\end{eqnarray}
which mimics the recent GUP proposal predicting the presence of both
a minimal length uncertainty and a maximal momentum uncertainty
through a doubly special relativity consideration
\cite{pedram,pedram3,pedram1,PRD}. There, the generalized momentum
has an upper bound proportional to $M_{Pl} c / \alpha_0$ where
$\alpha_0$ similar to $\beta_0$ is of the order of unity. However,
for our case, the generalized momentum $P$ has no upper bound and it
is not physically equivalent with aforementioned GUP. Therefore, the
idea of a maximum ``canonical'' momentum naturally arises from our
representation.

\section{GUP and the free particle}
In ordinary quantum mechanics, the free particle wave  function
$u_p(x)$ is defined as the eigenfunction of the momentum operator
$P_{op}$
\begin{eqnarray}\label{p}
P_{op}u_p(x)=p\,u_p(x),
\end{eqnarray}
where $p$ is the eigenvalue. The momentum operator has the following
representation in the quasiposition space
\begin{eqnarray}
P_{op}=\frac{\hbar}{i}\frac{\partial}{\partial x}.
\end{eqnarray}
So, from Eq.~(\ref{p}) we have
\begin{eqnarray}\label{p1}
\frac{\hbar}{i}\frac{\partial u_p(x)}{\partial x}=pu_p(x),
\end{eqnarray}
which has the following solution
\begin{eqnarray}\label{solp}
u_p(x)=\frac{1}{\sqrt{2\pi\hbar}}\exp\left({\frac{ip
x}{\hbar}}\right),
\end{eqnarray}
where the constant of integration is chosen to satisfy
\begin{eqnarray}\label{norm}
\int^{\infty}_{-\infty}u^{\ast}_p(x)u_p(x')\mathrm{d}p=\delta(x-x').
\end{eqnarray}
In GUP scenario, to find the momentum eigenfunction in the position
space, we write the momentum operator (\ref{x0p02}) as $P_{op}=
\tan\left(-i\hbar\sqrt{\beta}\partial_x\right)/\sqrt{\beta}$ which
results in the following eigenvalue equation
\begin{eqnarray}\label{p2}
\frac{\tan\left(-i\hbar\sqrt{\beta}\partial_x\right)}{\sqrt{\beta}}u_p(x)=pu_p(x).
\end{eqnarray}
Now, let us consider a class of solutions which satisfies
Eqs.~(\ref{p1}) and (\ref{p2}) at the same time, but with different
eigenvalues [$p\rightarrow p'$ in Eq.~(\ref{p1})], i.e.,
\begin{eqnarray}
u_p(x)=A(p)\exp\left({\frac{ip' x}{\hbar}}\right),
\end{eqnarray}
where $p'=f(p)$. Inserting this solution in Eq.~(\ref{p2}) results
in $\,\,\tan\left(\sqrt{\beta}p'\right)/\sqrt{\beta}=p\,\,$ or
\begin{eqnarray}\label{p'}
p'=\frac{1}{\sqrt{\beta}}\arctan\left(\sqrt{\beta}p\right),
\end{eqnarray}
so we have
\begin{eqnarray}\label{sai}
u_p(x)=A(p)\exp\left[\frac{i}{\hbar\sqrt{\beta}}\arctan\left(\sqrt{\beta}p\right)x\right].
\end{eqnarray}
To obtain $A(p)$, we demand that the momentum eigenfunction
satisfies Eq.~(23) of Ref.~\cite{Kempf} as the modified version of
(\ref{norm}) which results in
\begin{eqnarray}\label{Ap}
A(p)=\bigg[2\pi\hbar \left(1+\beta p^2\right)\bigg]^{-1/2}.
\end{eqnarray}
Thus, we finally obtain the momentum eigenfunctions as
\begin{eqnarray}
u_p(x)=\frac{1}{\sqrt{2\pi\hbar\left(1+\beta
p^2\right)}}\exp\left[\frac{i}{\hbar\sqrt{\beta}}\arctan\left(\sqrt{\beta}p\right)x\right],\hspace{.6cm}
\end{eqnarray}
which, to the first order agrees with the solution presented in
Ref.~\cite{106} i.e.
\begin{eqnarray}
u_p(x)=\left(\frac{1-\beta p^2}{2\pi\hbar}\right)^{1/2}\exp\left[\frac{i}{\hbar}\left(p-\frac{\beta}{3}p^3\right)x\right].
\end{eqnarray}
Note that this solution for $\beta\rightarrow0$ reduces to
(\ref{solp}) in order to satisfy the  correspondence principle.
Moreover, this result is similar to the position eigenvectors
obtained by KMM, where they used (\ref{k1}) and (\ref{k2}) subjected
to the deformed scalar product (\ref{scalar}). In fact, the factor
$1/\left(1+\beta p^2\right)$ in the definition of the scalar product
(\ref{scalar}) indeed appeared in the momentum-dependent
normalization coefficient of the momentum eigenfunctions, namely,
\begin{eqnarray}
|A(p)|^2\sim \frac{1}{1+\beta p^2}.
\end{eqnarray}

At this point, we can use the following theorem to check the
self-adjointness property of the position and momentum operators
\cite{Akhiezer,Bonneau}
\begin{theorem}
\emph{(von Neumann's theorem)} \label{Neumann} For an operator $A$
with deficiency indices $(n_+,n_-)$ there are three possibilities:

1. If $n_+=n_-=0$, then $A$ is self-adjoint (this is a necessary and
sufficient condition).

2. If $n_+=n_-=n\geq1$, then $A$ has infinitely many self-adjoint
extensions, parameterized by a unitary $n\times n$ matrix.

3. If $n_+\ne n_-$, then $A$ has no self-adjoint extension.
\end{theorem}

To use von Neumann's theorem, we have to find the wave functions
$\phi_{\pm}$ given by
\begin{eqnarray}
P^{\dagger}\phi_{\pm}(x)=\frac{\tan\left(-i\hbar\sqrt{\beta}\partial_x\right)}{\sqrt{\beta}}\phi_{\pm}(x)=\pm
i\lambda \phi_{\pm}(x).
\end{eqnarray}
So using Eq.~(\ref{sai}) we have
\begin{eqnarray}
\phi_{\pm}(x)=C_{\pm}\exp\left[\frac{\mp1}{\hbar\sqrt{\beta}}\tanh^{-1}\left(\sqrt{\beta}\lambda\right)x\right].
\end{eqnarray}
Since the operator $P$ is defined on the whole real axis where
$\phi_{\pm}$ diverge at $x\rightarrow\mp\infty$ and consequently are
not normalizable, none of the functions $\phi_{\pm}$ belong to the
Hilbert space ${\cal L}^2(\mathbb{R})$ and therefore the deficiency
indices are $(0,0)$. Hence, we conclude that the momentum operator
is indeed self-adjoint with the following domain
\begin{eqnarray}
{\cal D}(P)={\cal D}(P^{\dagger})=\left\{\phi\in{\cal
D}_{max}\left(\mathbb{R}\right)\right\},
\end{eqnarray}
where ${\cal D}_{max}$ denotes the maximal domain on which the
operator $P$ has a well defined action, i.e., ${\cal
D}_{max}(P)=\left\{\phi\in{\cal L}^2(\mathbb{R})\,:\,P\phi\in{\cal
L}^2(\mathbb{R})\right\}$. Using the same procedure for the position
operator $X$ on the finite interval, it is straightforward to check
that both $\phi_{\pm}(p)=C_{\pm}e^{\mp \lambda p}$ belong to ${\cal
L}^2(-\frac{1}{2}\pi\beta^{-1/2},\frac{1}{2}\pi\beta^{-1/2})$ and
the deficiency indices are $(1,1)$. Therefore, one concludes that
the position operator is no longer essentially self-adjoint but has
a one-parameter family of self-adjoint extensions which is in
agreement with the previous result.

\section{GUP and the harmonic oscillator}
In this section, we study the classical and quantum mechanical
solutions of the harmonic oscillator in the GUP framework and
present its semiclassical results. Moreover, we study the effects of
the minimal length uncertainty on the thermodynamic aspects of the
harmonic oscillator in both classical and quantum domains.

\subsection{Classical Description}
\begin{figure*}
\begin{center}
\begin{tabular}{ccc}
\includegraphics[width=5cm]{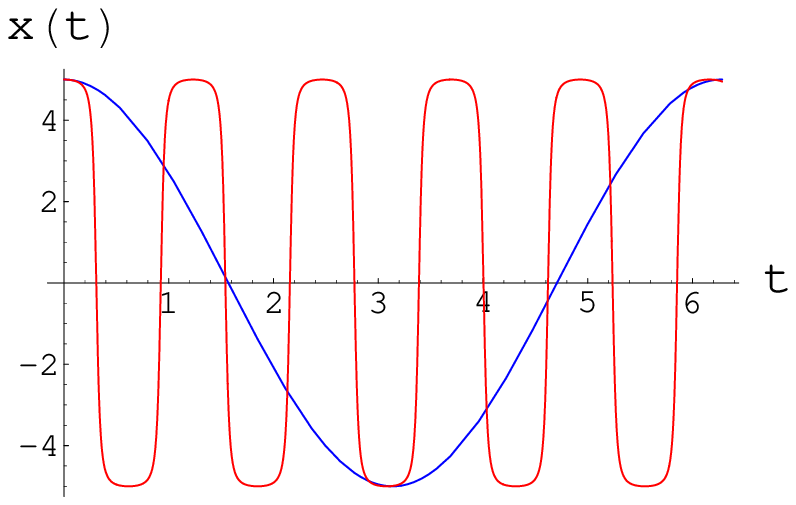}&\includegraphics[width=5cm]{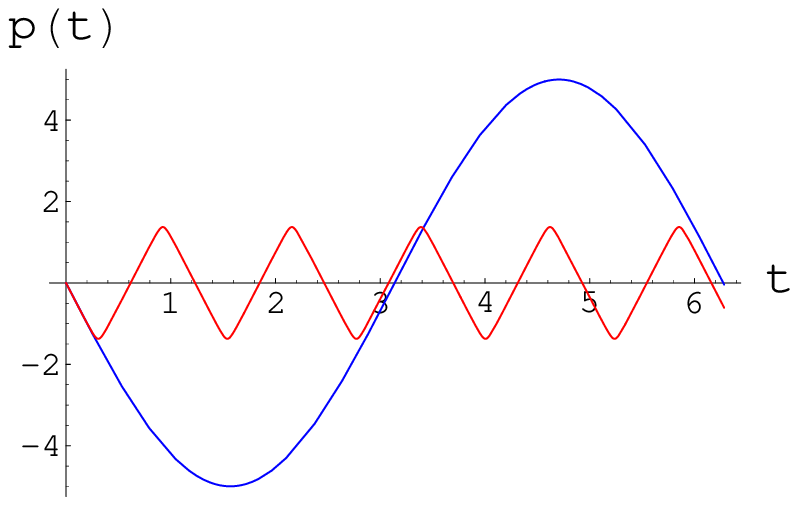}&\includegraphics[width=4cm]{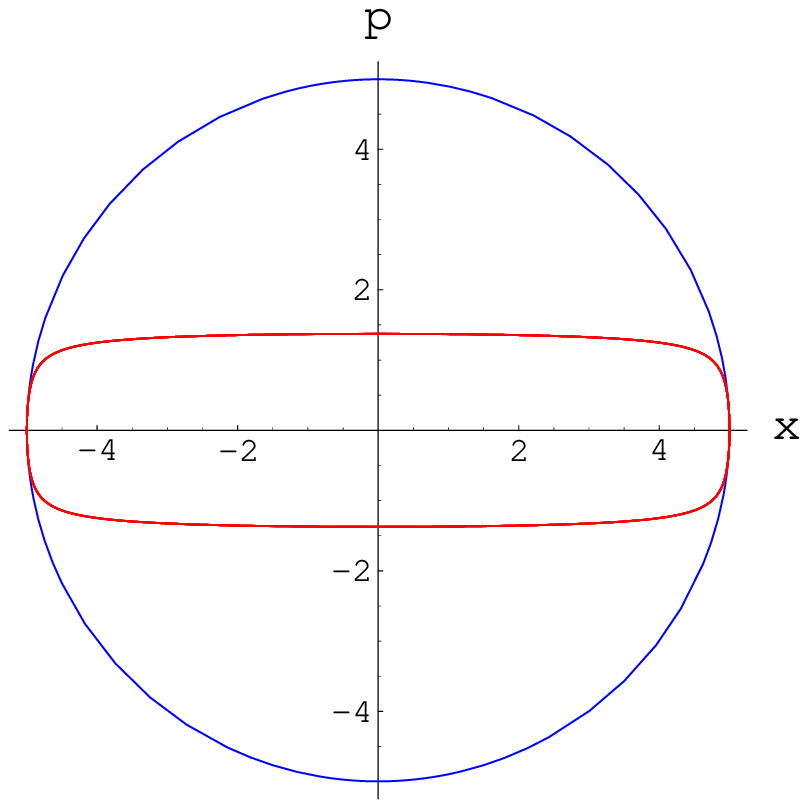}
\end{tabular}
\end{center}
\caption{\label{figx}The temporal behavior of $x$ and $p$, and the
phase space of the harmonic oscillator for $\beta=0$ (blue line) and
$\beta=1$ (red line). We set $m=\omega=1$ and $a=5$
($\bar{\omega}=\sqrt{26}$).}
\end{figure*}

Let us consider the Hamiltonian of a particle of mass $m$ confined
in a quadratic potential
\begin{eqnarray}\label{HSHO}
H^{(HO)}= \frac{\tan^2\left(\sqrt{\beta}p\right)}{2\beta
m}+\frac{1}{2}m\omega^2x^2,
\end{eqnarray}
which using the Hamiltonian equations results in
\begin{eqnarray}\label{HO1}
\dot{x}&=&\frac{\tan(\sqrt{\beta}p)\sec^2(\sqrt{\beta}p)}{\sqrt{\beta}m},\\
\dot{p}&=&-m\omega^2x.\label{HO2}
\end{eqnarray}
So, in the GUP formalism, the velocity $\dot{x}$ is not equal to
$p/m$, but it tends to $p/m$ as $\beta$ goes to zero. Using
Eqs.~(\ref{HO1}) and (\ref{HO2}) we obtain
\begin{eqnarray}\
\ddot{p}+\omega^2\frac{\tan(\sqrt{\beta}p)\sec^2(\sqrt{\beta}p)}{\sqrt{\beta}}=0.
\end{eqnarray}
If we set the initial conditions as $x(0)=a$ and $p(0)=0$, it is
straightforward to check that the above equation admits the
following solutions
\begin{eqnarray}
p(t)&=&\pm\frac{1}{\sqrt{\beta}}\arctan\left(\frac{\eta}{\sqrt{1+\left(1+\eta^2\right)\cot^2\left(
\sqrt{1+\eta^2}\omega t\right)}}\right),\hspace{.5cm}\\
x(t)&=&\mp\frac{a\sqrt{1+\eta^2}\cot\left( \sqrt{1+\eta^2}\omega
t\right)}{\sqrt{1+\left(1+\eta^2\right)\cot^2\left(
\sqrt{1+\eta^2}\omega t\right)}},
\end{eqnarray}
where $\eta=\sqrt{\beta}m\omega a$. So the actual frequency of the
harmonic oscillator in GUP scenario increases with respect to the
absence of GUP as $\bar{\omega}=\sqrt{1+\beta m^2\omega^2
a^2}\,\omega\geq\omega$. In fact, this frequency depends on GUP
parameter, particle's mass, and the initial position. Moreover, as
$\beta$ increases, the particle is often located at the end points
$\pm a$ and the accessible phase space decreases with respect to the
absence of GUP (see Fig.~\ref{figx}).

\subsection{Semiclassical description}
Before studying the corresponding generalized Schr\"odinger
equation, it is worthwhile to find the quantized energy spectrum
using the semiclassical scheme. The Wentzel-Kramers-Brillouin (WKB)
quantization rule, represented succinctly by the formula
\begin{eqnarray}
\oint
p\,\mathrm{d}q=\left(n+\frac{1}{2}\right)h,\hspace{2cm}n=0,1,\ldots,
\end{eqnarray}
allows us to find the approximate energy spectrum and in ordinary
quantum mechanics gives the exact results. Using Eq.~(\ref{HSHO}) we
find
\begin{eqnarray}
\oint
p\,\mathrm{d}x&=&\frac{2}{\sqrt{\beta}}\int_{-a}^{a}\arctan\left(\sqrt{\beta}
m\omega\sqrt{a^2-x^2}\right)\mathrm{d}x\nonumber\\
&=&2\pi\frac{\sqrt{1+\beta m^2\omega^2 a^2}-1}{\beta m\omega},
\end{eqnarray}
which results in the following semiclassical energy spectrum
\begin{eqnarray}
\hspace{-.6cm}E_n^{(SC)}&=&\frac{1}{2}m\omega^2a_n^2,\nonumber\\
\hspace{-.6cm}&=&-\frac{1}{8}\gamma\hbar\omega+\hbar\omega\left(n+\frac{1}{2}\right)\left(1+\gamma/2\right)+\frac{1}{2}\hbar\omega\gamma
n^2,\label{ESC}
\end{eqnarray}
where $\gamma=\beta m\hbar\omega$. As we have expected, $E_n^{(SC)}$
tends to $\hbar\omega\left(n+1/2\right)$ as $\beta$ goes to zero.
However, contrary to the ordinary formulation where $E_n^{(SC)}$ is
equal to the exact energy spectrum, it does not give the exact
spectrum in the GUP formalism. This is due to the fact that from the
Hamiltonian (\ref{HSHO}) we expect that the energy spectrum depends
on numerous powers of $\beta$, but $E_n^{(SC)}$ only represents a
linear dependence of the GUP parameter. In the next section, we show
this fact by a rigorous mathematical proof. However, it can be
considered as a good approximation which is related to the correct
quadratic dependence on the quantum number.

\subsection{Quantum Description}
For the case of the harmonic oscillator, because of the
quadratic form of the potential $V(x)=1/2\,m\omega^2x^2$, we obtain
a second-order differential equation in the momentum space, namely
\begin{eqnarray}\label{HamilSHO}
-\frac{\partial^2\phi(p)}{\partial
p^2}+\frac{\tan^2\left(\sqrt{\gamma}p\right)}{\gamma}\phi(p)=\bar{\epsilon}\,\phi(p),
\end{eqnarray}
where $p\rightarrow\sqrt{m\hbar\omega}\,p$,
$\gamma=m\hbar\omega\beta$, and
$\bar{\epsilon}=\frac{2E}{\hbar\omega}$. In terms  of the new
variable $z=\sqrt{\gamma}p$, we obtain
\begin{eqnarray}
\left[-\frac{\partial^2}{\partial
z^2}+\nu(\nu-1)\tan^2(z)-\epsilon(\nu)\right]\phi(z;\nu)=0,
\end{eqnarray}
where by definition
\begin{eqnarray}
\nu=\frac{1}{2}\left(1+\sqrt{1+\frac{4}{\gamma^2}}\right),\hspace{1cm}\epsilon(\nu)=\frac{\bar{\epsilon}}{\gamma},
\end{eqnarray}
and the boundary condition is
\begin{eqnarray}
\phi(z;\nu)\bigg|_{z=\pm\pi/2}=0.
\end{eqnarray}
The above differential equation is exactly solvable and the
eigenfunctions can be obtained in terms of Gauss hypergeometric
functions where we briefly present the solutions \cite{taseli}.

\begin{figure}
\begin{center}
\includegraphics[width=7cm]{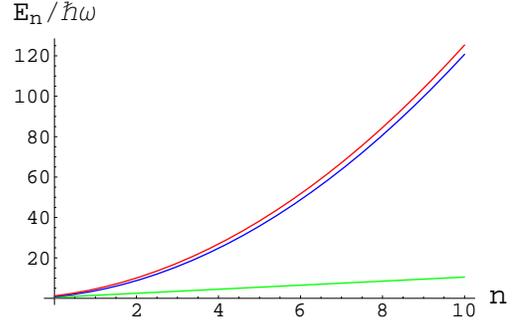}
\end{center}
\caption{\label{figE} Comparing $E_n/\hbar\omega$ (red line) and
$E_n^{(SC)}/\hbar\omega$ (blue line) for $\gamma=2$ with the
ordinary harmonic oscillator spectrum (green line).}
\end{figure}

To find the even parity states, let us use the substitution
$\xi=\sin^2(z)$ which leads to
\begin{eqnarray}
&&\xi(1-\xi)\frac{\partial^2\phi(\xi;\nu)}{\partial
\xi^2}+\left(\frac{1}{2}-\xi\right)\frac{\partial\phi(\xi;\nu)}{\partial
\xi}\nonumber\\
&&+\left[\Delta(\nu,\epsilon)-\frac{1}{4}\frac{\nu(\nu-1)}{1-\xi}\right]\phi(\xi;\nu)=0,
\end{eqnarray}
where $\Delta(\nu,\epsilon)=\frac{1}{4}[\nu(\nu-1)+\epsilon(\nu)]$.
Now, to get rid of the regular singularity of the last term we
search for the solution of the form
\begin{eqnarray}
\phi(\xi;\nu)=(1-\xi)^a\,Y(\xi;\nu),
\end{eqnarray}
where $a$ satisfy the algebraic equation
\begin{eqnarray}
a^2-\frac{1}{2}a-\frac{1}{4}\nu(\nu-1)=0.
\end{eqnarray}
So we obtain the Gauss hypergeometric equation for the variable
$Y(\xi;\nu)$
\begin{eqnarray}
\xi(1-\xi)Y''+\left(\frac{1}{2}-(\alpha+\beta+1)\xi\right)Y'-\alpha\beta
Y=0,
\end{eqnarray}
subjected to $\alpha+\beta=2a$ and
$\alpha\beta=a^2-\Delta(\nu,\epsilon)$. This equation admits two
independent solutions. However, the physically acceptable solution
which vanishes at the boundary $\lim_{\xi\rightarrow1}Y(\xi;\nu)=0$
is
\begin{eqnarray}
Y(\xi;\nu)={\cal A}(\nu)(1-\xi)^{\nu/2}\,{_2F_1}\left(\alpha,\beta;
\nu+ \frac{1}{2}; 1-\xi\right),
\end{eqnarray}
where ${\cal A}(\nu)$ is the normalization constant. The analyticity
and the convergence of the hypergeometric function for all
$\xi\in[0,1]$ results in
\begin{eqnarray}
\alpha\quad\mbox{or}\quad\beta=-k,\hspace{1cm}k=0,1,2,\ldots.
\end{eqnarray}
So we obtain the even parity eigenfunctions
\begin{eqnarray}
Y_{2k}(\xi;\nu)={\cal
A}_k(\nu)(1-\xi)^{\nu/2}\,{_2F_1}\left(-k,\nu+k; \nu+ \frac{1}{2};
1-\xi\right),\hspace{.6cm}
\end{eqnarray}
and the eigenvalues
\begin{eqnarray}\label{e1}
\epsilon_{2k}(\nu)=4k(\nu+k)+\nu,\hspace{1cm}k=0,1,2,\ldots.
\end{eqnarray}
Finally, in terms of the original variable $p$ we have
\begin{eqnarray}
\phi_{2k}(p;\gamma)&=&{\cal
A}_k(\nu)\left[\cos(\sqrt{\gamma}p)\right]^{\left(1+\sqrt{1+\frac{4}{\gamma^2}}\right)/2}\nonumber\\
&&\times{_2F_1}\left(-k,\nu+k; \nu+ \frac{1}{2};
\cos^2(\sqrt{\gamma}p)\right).\label{solSHO1}
\end{eqnarray}
To find the antisymmetric solutions let us define
\begin{eqnarray}
\phi(z;\nu)=\sin(z)\varphi(z;\nu),
\end{eqnarray}
where $\phi$ is an even function of $z$. By substitution of this
solutions in the original equation we have
\begin{eqnarray}
&&\left[-\frac{\partial^2}{\partial
z^2}-2\cot(x)\frac{\partial}{\partial
z}+\nu(\nu-1)\tan^2(z)+1-\epsilon(\nu)\right]\nonumber\\
&&\times\varphi(z;\nu)=0,
\end{eqnarray}
where by choosing $\xi=\sin^2(z)$ can be written as
\begin{eqnarray}
&\bigg[&\xi(1-\xi)\frac{\partial^2}{\partial
\xi^2}+\left(\frac{3}{2}-2\xi\right)\frac{\partial}{\partial
\xi}+\Delta(\nu,\epsilon)\nonumber\\
&&-\frac{1}{4}-\frac{1}{4}\frac{\nu(\nu-1)}{1-\xi}\bigg]\varphi(\xi;\nu)=0.\label{21}
\end{eqnarray}
Similar to the procedure for the even states let us define
\begin{eqnarray}
\phi(\xi;\nu)=(1-\xi)^{\nu/2}\,U(\xi;\nu),
\end{eqnarray}
which converts Eq.~(\ref{21}) to the Gauss hypergeometric equation
\begin{eqnarray}
\xi(1-\xi)U''+\left(\frac{3}{2}-(\bar{\alpha}+\bar{\beta}+1)\xi\right)U'-\bar{\alpha}\bar{\beta}
U=0,
\end{eqnarray}
where $\bar{\alpha}=\frac{1}{2}(\nu+1)-\sqrt{\Delta(\nu,\epsilon)}$
and $\bar{\beta}=\frac{1}{2}(\nu+1)+\sqrt{\Delta(\nu,\epsilon)}$. As
before we set $\bar{\alpha}=-k$ and find the eigenenergies
\begin{eqnarray}\label{e2}
\hspace{-.7cm}\epsilon_{2k+1}(\nu)=(2k+1)(2\nu+2k+1)+\nu,\hspace{.5cm}k=0,1,\ldots,
\end{eqnarray}
for the antisymmetric eigenfunctions
\begin{eqnarray}
U_{2k+1}(\xi;\nu)&=&{\cal
B}_k(\nu)\sqrt{\xi}(1-\xi)^{\nu/2}\nonumber\\
&&\times{_2F_1}\left(-k,\nu+k+1; \nu+ \frac{1}{2}; 1-\xi\right).
\end{eqnarray}
In terms of the original variables we have
\begin{eqnarray}\label{solSHO2}
\hspace{-.7cm}\phi_{2k+1}(p;\gamma)&=&{\cal
B}_k(\nu)\sin(\sqrt{\gamma}p)
\left[\cos(\sqrt{\gamma}p)\right]^{\left(1+\sqrt{1+\frac{4}{\gamma^2}}\right)/2}\nonumber
\\ \hspace{-.7cm}&&\times{_2F_1}\left(-k,\nu+k+1; \nu+ \frac{1}{2};
\cos^2(\sqrt{\gamma}p)\right).
\end{eqnarray}
Note that we can combine Eqs.~(\ref{e1}) and (\ref{e2}) in a single
formula to express the full spectrum, namely
$\epsilon_{n}(\nu)=n(2\nu+n)+\nu$ for $n=0,1,2,\ldots$ or
\begin{eqnarray}
\hspace{-.7cm}E_n(\gamma)=\hbar\omega\left(n+\frac{1}{2}\right)
\left(\sqrt{1+\gamma^2/4}+\gamma/2\right)+\frac{1}{2}\hbar\omega\gamma
n^2,\label{Eexact}
\end{eqnarray}
in terms of $\gamma$. So, as we have expected, this result exactly
coincides with the spectrum of the harmonic oscillator in the
formalism proposed by Kempf, Mangano, and Mann. In Fig.~\ref{figE},
we have depicted the energy spectrum of the harmonic oscillator in
GUP framework (\ref{Eexact}), its semiclassical approximation
(\ref{ESC}), and its spectrum in ordinary quantum mechanics. The
efficiency of the semiclassical solution is manifest in the figure.
In fact, to first order of the GUP parameter, $E_n$ is equal to
$E_n^{(SC)}$ up to a positive constant, namely
\begin{eqnarray}
E_n\simeq E_n^{(SC)}+\frac{1}{8}\gamma\hbar\omega.
\end{eqnarray}

To check the self-adjointness property of $H^{(HO)}$, it is natural
to present the sesquilinear form for $\psi$ and $\phi$ as
\begin{eqnarray}
&2&iB(\psi,\phi)=\langle H\psi|\phi\rangle-\langle
\psi|H|\phi\rangle,\nonumber\\
&=&\int_{-\frac{\pi}{2\sqrt{\beta}}}^{+\frac{\pi}{2\sqrt{\beta}}}\mathrm{d}p\,\left(H\psi(p)\right)^{*}\phi(p)-
\int_{-\frac{\pi}{2\sqrt{\beta}}}^{+\frac{\pi}{2\sqrt{\beta}}}\mathrm{d}p\,\psi^{*}(p)H\phi(p),\nonumber\\
&=&-\frac{1}{2}m\omega^2\hbar^2\left[\int_{-\frac{\pi}{2\sqrt{\beta}}}^{+\frac{\pi}{2\sqrt{\beta}}}\mathrm{d}p\,\psi''(p)^{*}\phi(p)-
\int_{-\frac{\pi}{2\sqrt{\beta}}}^{+\frac{\pi}{2\sqrt{\beta}}}\mathrm{d}p\,\psi^{*}(p)\phi''(p)\right],\nonumber\\
&=&\frac{1}{2}m\omega^2\hbar^2\Bigg[\psi^*(p)\phi'(p)\Bigg|_{p=\frac{\pi}{2\sqrt{\beta}}}\hspace{-.75cm}
-\psi'^*(p)\phi(p)\Bigg|_{p=\frac{\pi}{2\sqrt{\beta}}}\nonumber\\&&
-\psi^*(p)\phi'(p)\Bigg|_{p=\frac{-\pi}{2\sqrt{\beta}}}\hspace{-.75cm}+\psi'^*(p)\phi(p)\Bigg|_{p=\frac{-\pi}{2\sqrt{\beta}}}\Bigg].
\end{eqnarray}
On the other hand, using the explicit form of the solutions
(\ref{solSHO1}) and (\ref{solSHO2}), it is straightforward to check
that the first derivative of the solutions as well as $\phi(p;\nu)$
vanishes at the boundaries i.e.
\begin{eqnarray}
\phi'(p;\nu)\bigg|_{p=\frac{\pm\pi}{2\sqrt{\beta}}}=0.
\end{eqnarray}
Therefore, $\psi^*(p)$ and $\psi'^*(p)$ can take arbitrary values at
the boundaries. This means that the domain of the adjoint of $H$ is
larger than that of $H$, so the Hamiltonian is symmetric but not
self-adjoint. The domains are
\begin{eqnarray}
\hspace{-2cm}{\cal D}(H)&=&\bigg\{\phi\in{\cal
D}_{max}\left(\frac{-\pi}{2\sqrt{\beta}},\frac{+\pi}{2\sqrt{\beta}}\right)\,;
\phi\left(\frac{+\pi}{2\sqrt{\beta}}\right)\nonumber\\
&=&\phi\left(\frac{-\pi}{2\sqrt{\beta}}\right)
=\phi'\left(\frac{+\pi}{2\sqrt{\beta}}\right)=\phi'\left(\frac{-\pi}{2\sqrt{\beta}}\right)=0\bigg\},
\end{eqnarray}
\begin{eqnarray}
{\cal D}(H^{\dagger})=\bigg\{\psi\in{\cal
D}_{max}\left(\frac{-\pi}{2\sqrt{\beta}},\frac{+\pi}{2\sqrt{\beta}}\right);\nonumber\\
\quad\quad\mbox{no\,\,other\,\,restriction\,\,on }\psi\bigg\}.
\end{eqnarray}
Note that this result is not surprising because even the Hamiltonian
of the one-dimensional particle in a box is not a truly self-adjoint
operator as well \cite{Bonneau}.

\subsection{Classical partition function}
\begin{figure}
\begin{center}
\includegraphics[width=7cm]{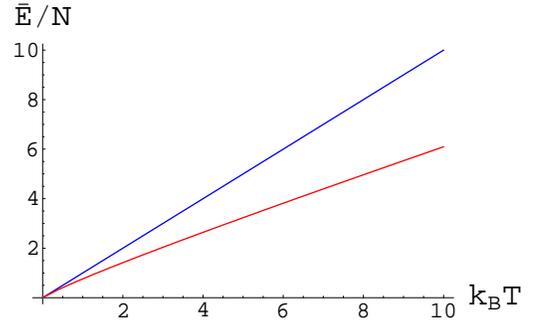}
\end{center}
\caption{\label{figEbar} The classical mean energy of $N$ harmonic
oscillators in thermal equilibrium versus temperature for $\beta=1$
(red line) and $\beta=0$ (blue line). We set $m=1$.}
\end{figure}

In statistical mechanics, the canonical partition function of $N$
identical, one-dimensional oscillators which encodes the statistical
properties of a thermodynamic system can be written in the classical
domain as
\begin{eqnarray}
Z(b)&=&\frac{1}{N! h^{N}} \int \, \exp\left[-b H(p_1 \cdots p_N, x_1
\cdots x_N)\right]\nonumber\\ &&\times\, \mathrm{d}p_1 \cdots
\mathrm{d}p_N \, \mathrm{d}x_1 \cdots \mathrm{d}x_N,
\end{eqnarray}
where $b\equiv1/k_BT$, $k_B$ denotes Boltzmann's constant and $T$ is
the temperature. For $N$ noninteracting oscillators, the total
partition function can be obtained from the single-particle
partition function as
\begin{eqnarray}
Z(b)&=&\frac{1}{N!h^{N}}\left[ \int_{-\infty}^{+\infty}
\exp\left[-\frac{1}{2}b m\omega^2x^2 \right]\mathrm{d}x \right]^N
\nonumber\\
&&\times\,\left[
\int_{-\frac{\pi}{2\sqrt{\beta}}}^{+\frac{\pi}{2\sqrt{\beta}}}\exp\left[-\frac{b}{2\beta
m}\tan^2(\sqrt{\beta}p) \right]\mathrm{d}p  \right]^N,
\end{eqnarray}
\begin{eqnarray}
&=&\frac{1}{N!h^{N}}\left(\frac{2\pi
k_BT}{m\omega^2}\right)^{N/2}\nonumber\\
&&\times\,\left(\frac{\pi \exp\left(\frac{1}{2\beta m
k_BT}\right)\mathrm{erfc}\left(\frac{1}{\sqrt{2\beta
mk_BT}}\right)}{\sqrt{\beta}}\right)^N,
\end{eqnarray}
where $\mathrm{erfc}(x)$ is the complementary error function. Using
the asymptotic expansion of the complementary error function for
large $x$, namely
\begin{eqnarray}
\mathrm{erfc}(x) = \frac{e^{-x^2}}{x\sqrt{\pi}}\left
[1+\sum_{n=1}^\infty (-1)^n
\frac{1\cdot3\cdot5\cdots(2n-1)}{(2x^2)^n}\right],
\end{eqnarray}
we can write the partition function in terms of powers of $k_BT$ as
\begin{eqnarray}
Z(b)=\frac{1}{N!}\left(\frac{
k_BT}{\hbar\omega}\right)^{N}\left(1+\sum_{n=1}^{\infty}(2n-1)!!\left(-\beta
mk_BT\right)^n\right)^N,\hspace{.6cm}
\end{eqnarray}
where for $\beta\rightarrow0$ reduces to the ordinary partition
function. Also, the classical mean energy of the system is given by
\begin{eqnarray}
\overline{E}_C&=&-\frac{\partial}{\partial b}\ln
Z\nonumber\\
&=&N\left(\frac{k_BT}{2}+\sqrt{\frac{k_BT}{2\pi\beta
m}}\frac{\exp\left(-1/2\beta m
k_BT\right)}{\mathrm{erfc}\left(\sqrt{1/2\beta
mk_BT}\right)}-\frac{1}{2\beta m}\right),\hspace{.6cm}\label{Z0} \\
&=&N\left(k_BT+\frac{\sum_{n=1}^{\infty}n(2n-1)!!\left(-\beta
m\right)^n(k_BT)^{n+1}}{1+\sum_{n=1}^{\infty}(2n-1)!!\left(-\beta
mk_BT\right)^n}\right),\label{Z}
\end{eqnarray}
which goes to $Nk_BT$ for $\beta\rightarrow0$. Therefore, as
indicated in Fig.~\ref{figEbar}, in the presence of GUP, the mean
energy decreases with respect to $\beta=0$. The reason for the
reduction of mean energy with respect to $\beta=0$ is a consequence
of the reduction of phase space volume (surface) due to possible
definition of a rescaled $\hbar$. In fact the volume of the
fundamental cell increases in the presence of the minimal length
uncertainty relation and the number of degrees of freedom reduces
consequently. Moreover, it modifies the Helmholtz free energy
$A=-k_BT\ln Z$ and the entropy $S=k_B\ln Z+\overline{E}/T$ as well.
The above equation shows that the equipartition theorem fails in the
GUP scenario. Although the averaged potential satisfies the
equipartition theorem i.e. $\langle 1/2 m \omega^2
x^2\rangle=k_BT/2$, the kinetic part yields the smaller value
$\langle K\rangle<k_BT/2$ [see Eq.~(\ref{Z})]. Similarly, the heat
capacity at constant volume which is proportional to $\frac{\partial
\overline{E}}{\partial T}$, decreases with respect to the absence of
GUP, namely
\begin{eqnarray}
C_V^{\beta\ne0}<C_V^{\beta=0}.
\end{eqnarray}
Note that, for the case of the ideal gas we can write the partition
function as
\begin{eqnarray}
\hspace{-1cm}Z(b)=\frac{V^N}{N!h^{N}}\left(\frac{\pi
\exp\left(1/2\beta m k_BT\right)\mathrm{erfc}\left(\sqrt{1/2\beta
mk_BT}\right)}{\sqrt{\beta}}\right)^N,
\end{eqnarray}
so, using the definition of pressure $P=\frac{1}{b}\frac{\partial
\ln Z}{\partial V}$, we recover the ordinary ideal gas equation of
state $PV=Nk_BT$. However, the corresponding heat capacity will be
modified  as mentioned above.

\subsection{Quantum partition function}
\begin{figure}
\begin{center}
\includegraphics[width=7cm]{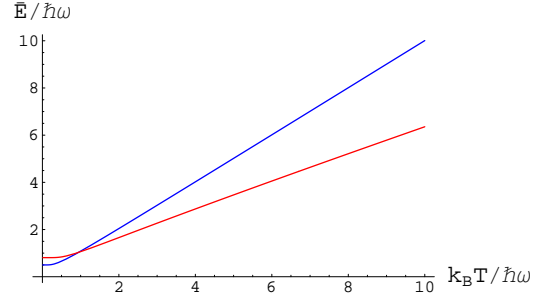}
\end{center}
\caption{\label{figEQ} The quantum mechanical mean energy of the
harmonic oscillator $E_n/\hbar\omega$ versus $k_BT/\hbar\omega$ for
$\gamma=1$ (red line) and $\gamma=0$ (blue line).}
\end{figure}

\begin{figure}
\begin{center}
\includegraphics[width=7cm]{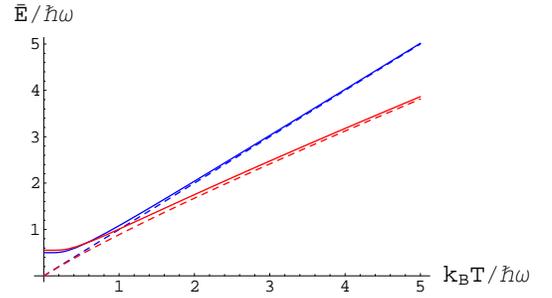}
\end{center}
\caption{\label{figECompare} The classical (dashed line) and quantum
mechanical (solid line) mean energy of the harmonic oscillator for
$\gamma=0.2$ (red line) and $\gamma=0$ (blue line).}
\end{figure}

In the quantum statistical mechanics, the partition function for a
single oscillator is given by
\begin{eqnarray}
Z(b)=\sum_{n=0}^{\infty}\exp\left(-bE_n\right),
\end{eqnarray}
where the energy eigenvalues are defined in Eq.~(\ref{Eexact}). Now,
\begin{eqnarray}
Z(b;\gamma)&=&e^{-(1/2)b\hbar\omega\left(\sqrt{1+\gamma^2/4}+\gamma/2\right)}\nonumber\\
&&\hspace{-2cm}\times\,
\sum_{n=0}^{\infty}\exp\left[-b\hbar\omega\left(\left(\sqrt{1+\gamma^2/4}+\gamma/2\right)n+\frac{1}{2}\gamma
n^2\right)\right],\\
&=&e^{-(1/2)b\hbar\omega\left(\sqrt{1+\gamma^2/4}+\gamma/2\right)}P(b;\gamma),
\end{eqnarray}
where we defined $P(b;\gamma) \equiv \sum_{n=0}^{\infty}
\exp\left[-b\hbar\omega \left(\left(\sqrt{1+\gamma^2/4}
+\gamma/2\right)n+\frac{1}{2}\gamma n^2\right)\right]$. So we have
$P(b;0)=\frac{1}{1-\exp(-b\hbar\omega)}$ and
$Z(b;0)=\frac{\exp(-(1/2)b\hbar\omega)}{1-\exp(-b\hbar\omega)}$.
Also, the mean energy of the oscillator is given by
\begin{eqnarray}
\overline{E}&=&-\frac{\partial}{\partial b}\ln
Z=\frac{1}{2}\hbar\omega\left(\sqrt{1+\gamma^2/4}+\gamma/2\right)-\frac{P'(b;\gamma)}{P(b;\gamma)},\hspace{1cm}\\
&=&
\hbar\omega\left(\sqrt{1+\gamma^2/4}+\gamma/2\right)\nonumber\\
&&\hspace{-2.5cm}\times\left(\frac{1}{2}+\frac{\sum_{n=0}^{\infty}\left(n+\frac{n^2}{1+\sqrt{1+4/\gamma^2}}\right)e^{-\frac{\hbar\omega}{k_BT}
\left(\left(\sqrt{1+\gamma^2/4}+\gamma/2\right)n+\frac{1}{2}\gamma
n^2\right)}}{\sum_{n=0}^{\infty}e^{-\frac{\hbar\omega}{k_BT}
\left(\left(\sqrt{1+\gamma^2/4}+\gamma/2\right)n+\frac{1}{2}\gamma
n^2\right)}}\right),\label{Z2}
\end{eqnarray}
where prime denotes the derivative with respect to $b$. The mean
energy of the harmonic oscillator in the quantum domain and in the
GUP formalism is depicted in Fig.~\ref{figEQ} which shows a modified
minimum value in the low temperature limit
\begin{eqnarray}
\overline{E}\simeq\frac{1}{2}\hbar\omega\left(\sqrt{1+\gamma^2/4}+\gamma/2\right).
\end{eqnarray}
To compare the classical and quantum results in the high-temperature
limit, we can write Eq.~(\ref{Z0}) as
\begin{eqnarray}\label{EClass}
\hspace{-.5cm}\frac{\overline{E}_C}{N\hbar\omega}=\frac{1}{2}\left(\frac{k_BT}{\hbar\omega}+\sqrt{\frac{k_BT/\hbar\omega}{2\pi\gamma}}\frac{\exp\left(\frac{-1/2\gamma}{
k_BT/\hbar\omega}\right)}{\mathrm{erfc}\left(\sqrt{\frac{1/2\gamma}{
k_BT/\hbar\omega}}\right)}-\frac{1}{\gamma}\right).\hspace{.6cm}
\end{eqnarray}
In Fig.~\ref{figECompare}, the classical (\ref{EClass}) and the
quantum mechanical (\ref{Z2}) mean energy of the harmonic oscillator
for $\gamma=0.2$ is depicted and compared with the ordinary
thermodynamic results.

\section{GUP and the potentials with sharp boundaries}
In the GUP scenario, we cannot measure the position of a particle
with an uncertainty less than $(\Delta X)_{min}$. So, in principle,
it is not possible to properly define the potentials with infinitely
sharp boundaries (It is well known that such sharp boundaries cannot
be also defined in theories with space-time uncertainty
\cite{h1,h2}). Indeed, the position of the boundaries can be only
determined within this uncertainty. However, one may argue that in a
first-step analysis, the assumption of sharp boundaries would be an
acceptable approximation. But the validity of this approximation
requires that the uncertainty in the energy spectrum due to the
boundaries' position uncertainty to be much smaller than the GUP
energy correction.

To investigate this point, we study the problem of a particle in a
box as an example of a potential with infinitely sharp boundaries
and compare both energy corrections. So, let us consider a particle
with mass $m$ confined in an infinite one-dimensional box with
length $L$
\begin{eqnarray}\label{pot}
V(x)=\left\{
\begin{array}{ll}
0 \hspace{1cm} \,\,0<x<L,\\ \infty \hspace{1cm}\mbox{elsewhere} .
\end{array}
\right.
\end{eqnarray}
The corresponding eigenfunctions should satisfy the following
generalized Schr\"odinger equation
\begin{eqnarray}
&-&\frac{\hbar^2}{2m}\frac{\partial^2\psi_n(x)}{\partial
x^2}+\sum_{j=3}^\infty \alpha_j \hbar^{2(j-1)}\beta^{j-2}
\frac{\partial^{2(j-1)}\psi(x)}{\partial
x^{2(j-1)}}\nonumber\\
&&=E_n\,\psi_n(x),\label{H2}
\end{eqnarray}
for $0<x<L$ and they also meet the boundary  conditions
$\psi_n(0)=\psi_n(L)=0$. Because of the boundary conditions, the
eigenfunctions do not change with respect to the absence of the GUP
($\beta=0$) \cite{106}. This fact leads us to consider the following
additional condition for the eigenfunctions
\begin{eqnarray}\label{H3}
-\frac{\hbar^2}{2m}\frac{\partial^2\psi_n(x)}{
\partial x^2}=\varepsilon_n\psi_n(x),\hspace{1cm}0<x<L,
\end{eqnarray}
where $\varepsilon_n=\frac{ n^2\pi^2\hbar^2}{ 2mL^2}$. If this
condition is also satisfied, we can write the second term in
Eq.~(\ref{H2}) in terms of $\psi_n(x)$ i.e.
\begin{eqnarray}
\frac{\partial^{2(j-1)} \psi_n(x)}{\partial
x^{2(j-1)}}&=&\frac{-2m\varepsilon_n}{\hbar^{2}}\frac{\partial^{2(j-2)}\psi_n(x)}{\partial
x^{2(j-2)}}=\cdots\nonumber\\
&=&\left(\frac{-2m\varepsilon_n}{\hbar^{2}}\right)^{j-1}\psi_n(x).
\end{eqnarray}
So, we have
\begin{eqnarray}
&-&\frac{\hbar^2}{2m}\frac{\partial^2\psi_n(x)}{\partial
x^2}+\sum_{j=3}^\infty \alpha_j \hbar^{2(j-1)}\beta^{j-2}
\frac{\partial^{2(j-1)}\psi_n(x)}{\partial x^{2(j-1)}}\nonumber\\
&&= \left(\varepsilon_n+\sum_{j=3}^\infty |\alpha_j|
\beta^{j-2}(2m)^{j-1}\varepsilon_n^{j-1}\right)\psi_n(x).\label{H4}
\end{eqnarray}
Now, comparing Eqs.~(\ref{H2}) and (\ref{H4}) shows that
\begin{eqnarray}\label{EBox}
E_n&=&\frac{\tan^2\left(\sqrt{2m\beta\varepsilon_n}\right)}{2m\beta}, \\
&=&\varepsilon_n+\frac{4}{3}\beta m
\varepsilon_n^{2}+\frac{68}{45}\beta^2 m^2
\varepsilon_n^{3}+\frac{496}{315}\beta^3 m^3
\varepsilon_n^{4}+\cdots,\nonumber\\
 &=&\frac{\hbar^2}{mL^2}\bigg[\frac{n^2\pi^2}{2}+\frac{n^4\pi^4}{3}\left(\frac{(\Delta
 X)_{min}}{L}\right)^2\nonumber\\ &&+\frac{17n^6\pi^6}{90}\left(\frac{(\Delta
 X)_{min}}{L}\right)^3+\cdots\bigg].
\end{eqnarray}
This GUP corrected energy spectrum can be also obtained using the
Wilson-Sommerfeld quantization rule given by
\begin{eqnarray}
\oint p\,\mathrm{d}q=nh,\hspace{1cm}n=1,2,\ldots,
\end{eqnarray}
with two conjugate variables $p$ and $q$ and the integer $n$. Since
the potential is constant (zero) inside the box, we have
\begin{eqnarray}
\oint
p\,\mathrm{d}x=\frac{2L}{\sqrt{\beta}}\arctan\left(\sqrt{2\beta
mE_n}\right).
\end{eqnarray}
So the semiclassical spectrum is
\begin{eqnarray}
E_n^{(SC)}=\frac{\tan^2\left(\sqrt{\beta}n\pi\hbar/L\right)}{2m\beta},
\end{eqnarray}
which exactly coincides with the quantum mechanical spectrum given
by Eq.~(\ref{EBox}). These results show that the GUP energy
correction is of order of $\left(\frac{(\Delta
 X)_{min}}{L}\right)^2$.

Now let us find the energy correction due to the uncertainty in the
position of the well's walls
\begin{eqnarray}
\Delta E_n\simeq \left|\frac{d\varepsilon_n}{dL}\right|(\Delta
 X)_{min}=\frac{n^2\pi^2\hbar^2}{mL^2}\left(\frac{(\Delta
 X)_{min}}{L}\right),\hspace{.5cm}
\end{eqnarray}
which is  first order in $(\Delta
 X)_{min}/L$. Therefore, the GUP energy correction is much smaller
than $\Delta E_n$ and cannot be detected in the presence of the
minimal length. This result confirms that the particle in a box
potential cannot be defined in the GUP framework as in ordinary
quantum mechanics. This conclusion can be also generalized to other
potentials with infinitely sharp boundaries.

\section{Conclusions}
In this paper, we proposed a nonperturbative gravitational quantum
mechanics in agreement with the existence of a minimal length
uncertainty relation. In this formalism the generalized Hamiltonian
takes the form $H=\tan^2\left(\sqrt{\beta}p\right)/(2\beta m)+V(x)$,
where $x$ and $p$ are the ordinary position and momentum operators.
We showed that this approach is equivalent with KMM representation
and we found the corresponding canonical transformation. This
representation has some advantages: First, it modifies only the
kinetic part (momentum operator) and the potential term (position
operator) remains unchanged. Second, this formalism is compatible
with perturbative schemes. Third, this representation predicts the
existence of a maximal canonical momentum proportional to $M_{Pl}
c/\sqrt{\beta_0}$. Because of the universality of the GUP effects,
this formalism can potentially be tested in various quantum
mechanical systems, of which we have studied just a few cases.

We thoroughly studied the case of the harmonic oscillator in
classical and quantum domains. In the classical domain, we found the
trajectory of the oscillating particle and showed that the GUP
modified frequency of the oscillator depends on mass, initial
position and the GUP parameter. Also, for large $\beta$ the particle
is often located around the end points. In the quantum domain, we
obtained the exact energy eigenvalues and the eigenfunctions and
showed that they are in agreement with those obtained in
Ref.~\cite{Kempf}. Moreover, the quadratic dependence of the energy
spectrum on the state number is confirmed using the semiclassical
approximation. To address the effects of the generalized uncertainty
principle on the thermodynamic properties of the harmonic
oscillator, we found the partition functions and the mean energies
in both classical and quantum limits. We showed that, in the
presence of the GUP and at the fixed temperature, the mean energy
and the heat capacity of the oscillator reduce in comparison with
those of the ordinary classical and quantum mechanics.

Also, we have indicated that $X$ and $H^{(HO)}$ are merely
symmetric, but $P$ is a truly self-adjoint operator. Note that these
results for $X$ and $P$ agree with those of KMM representation
\cite{Kempf}. However, the difference is that in this representation
all these operators are formally self-adjoint, i.e., $A=A^{\dagger}$
($A\in\{X,P,H^{(HO)}\}$), but ${\cal D}(A)\ne{\cal D}(A^{\dagger})$
for $A\in\{X,H^{(HO)}\}$ and ${\cal D}(P)={\cal D}(P^{\dagger})$. On
the other hand, in KMM representation only $P$ is formally and truly
self-adjoint. The problems with the potentials with sharp boundaries
are finally discussed. We showed that for this type of potentials,
the GUP energy correction is much smaller than the uncertainty in
the energy spectrum due to the boundaries' position uncertainty.

\acknowledgments I am very grateful to Kourosh Nozari, Rajesh
Parwani and Achim Kempf for fruitful discussions and suggestions and
for a critical reading of the manuscript. I would like to thank the
referees for giving such constructive comments which considerably
improved the quality of the paper.

\end{document}